\documentstyle[epsfig,12pt]{article}
\newcommand{\Z}{{\sf Z \!\!\! Z}}

\makeatletter
\@addtoreset{equation}{section}
\makeatother
 
\title{Confinement in the Deconfined Phase: A numerical study with
       a cluster algorithm
\footnote{This work is supported in part by funds provided by the U.S.
Department of Energy (D.O.E.) under cooperative research agreement
DE-FC02-94ER40818.}}
\author{K. Holland \\ \\
Center for Theoretical Physics, \\
Laboratory for Nuclear Science, and Department of Physics \\
Massachusetts Institute of Technology (MIT) \\
Cambridge, Massachusetts 02139, U.S.A. \\ \\
{\tt holland@ctp.mit.edu} \\ \\
MIT Preprint, CTP 2832 \\ \\}
 
\begin{document}
\maketitle
\begin{abstract} \normalsize
 
We have previously found analytically a very unusual and unexpected form of
confinement in $SU(3)$ Yang-Mills theory. This confinement occurs in the 
deconfined phase of the theory. The free energy of a single static test quark
diverges, even though it is contained in deconfined bulk phase and there is 
no QCD string present. This phenomenon occurs in cylindrical volumes with a certain
choice of spatial boundary conditions. We examine numerically an effective model for 
the Yang-Mills theory and, using a cluster algorithm, we observe this
unusual confinement. We also find a new way to determine the interface
tension of domain walls separating distinct bulk phases.

\end{abstract}
 
\maketitle
 
\newpage

\section{Introduction}

In this paper, we investigate numerically an analytical result
discussed in a previous paper \cite{Hol97}, where we derived the finite 
volume dependence of the Polyakov loop in cylinders with particular boundary
conditions. The analytical result tells us that, even in the deconfined 
phase, an external static quark is still confined. Confinement of quarks is 
usually described by a string tension $\sigma$, which is the energy cost per 
unit length of the QCD string connecting the fundamental charges. This 
unusual confinement is due to the multiple degenerate deconfined phases and 
has an effective ``string tension'' $\sigma^{\prime}$, which depends on the 
energy cost of a domain wall separating two deconfined phases. In this paper, 
we observe numerically this unusual confinement in an effective model for the 
Yang-Mills theory. Using a cluster algorithm, we can accurately determine the 
energy cost of these domain walls. In addition, we derive a new analytical 
result for the finite volume behavior of the Polyakov loop near the 
deconfinement phase transition, which we also examine numerically with a 
cluster algorithm. In Section 2, we present the analytical calculations, in 
Section 3, we give the details of the numerical work and in Section 4, we 
briefly draw our conclusions.

\section{Analytical dilute interface gas calculations}
$SU(3)$ Yang-Mills theory at non-zero temperature lives in
four-dimensional Euclidean space-time, with gauge fields 
$A_\mu(\vec{x},t) = ieA^a_\mu(\vec{x},t)\lambda^a$. The non-zero temperature is
$T = 1/\beta$, where $\beta$ is the extent in the periodic time direction, i.e.
$A_\mu(\vec{x},t + \beta) = A_\mu(\vec{x},t)$. The action $S[A_\mu] = 
\int^\beta_0 dt \int d^3 x (1/2e^2) \mbox{Tr} F_{\mu \nu} F_{\mu \nu}$, is 
constructed from the field strength $F_{\mu \nu} = 
\partial_\mu A_\nu - \partial_\nu A_\mu + [A_\mu, A_\nu]$. The action is 
invariant under gauge transformations $A^{\prime}_\mu = g^\dagger
(A_\mu + \partial_\mu)g$, where the $SU(3)$ matrices $g(\vec{x},t)$ are also
periodic in the time direction. From our point of view, the most important
field in the theory is the Polyakov loop, 
\begin{equation}
\Phi(\vec{x}) = \mbox{Tr}[{\cal P} \exp \int^\beta_0 dt A_4(\vec{x},t)],
\end{equation}
which is complex-valued and is built from the Euclidean time component of the 
gauge field. The Polyakov loop measures the response of the gauge field to 
the presence of a static test quark. Quantitatively, the free energy $F$ of 
the test quark is given by the expectation value $\langle \Phi \rangle = 
\exp(- \beta F)$. If the quark is confined, then $F$ diverges and $\langle 
\Phi \rangle = 0$, while if the quark is deconfined, $F$ is finite and 
$\langle \Phi \rangle \ne 0$. The Polyakov loop transforms non-trivially 
under a global symmetry of the action $S$. If the gauge transformations are 
not quite periodic in the time direction, but are instead twisted, i.e.
\begin{equation}
g(\vec{x},t + \beta) = g(\vec{x},t) z ~, ~ z \in \Z(3) = 
\{\exp(2 \pi i n/3), n = 1,2,3\},
\end{equation}
the action is invariant but the Polyakov loop changes into $\Phi'(\vec{x})
= \Phi(\vec{x}) z$. If $\langle \Phi \rangle = 0$, then the Polyakov loop
value does not break this global symmetry --- zero times $z$ is still zero.
However, if $\langle \Phi \rangle \ne 0$, this global symmetry is 
spontaneously broken. The Polyakov loop is the order parameter for this global 
symmetry --- the $\Z(3)$ center symmetry. If the symmetry is intact, the quark is 
confined; if it is broken, the quark is deconfined. In fact, we know that the 
symmetry is spontaneously broken at high temperatures \cite{Mcl81}. 

We study finite volumes to see whether or not the global symmetry is 
spontaneously broken in the infinite volume limit. In finite volumes, the
system is sensitive to the spatial boundary conditions. For example, we could 
choose all spatial directions to obey periodic
boundary conditions. However, $\langle \Phi \rangle$ would then always be
zero, independent of temperature. This is because the Polyakov loop represents
a static charge and Gauss' law implies that one can't have a charge contained in 
a torus \cite{Hil83}. We instead apply periodic boundary conditions in only
two spatial directions, call them $x$ and $y$. In the third direction $z$ we 
use what are called charge conjugate or $C$-periodic boundary conditions
\cite{Kro91}. If a field is $C$-periodic, it is replaced by its charge 
conjugate when translated by one period. For example, $C$-periodicity for 
gauge fields with period $L_z$ means that
\begin{equation}
A_\mu(\vec{x} + L_z \vec{e}_z,t) = A_\mu(\vec{x},t)^*,
\end{equation}
where $*$ denotes complex conjugation. Being constructed from the gauge 
field, the Polyakov loop satisfies $C$-periodicity in the $z$-direction and 
we ask that the gauge transformations also satisfy this boundary 
condition. Note that charge conjugation is a symmetry of the Yang-Mills 
action. It is a useful boundary condition to apply because the center 
electric flux coming out of a quark can escape and end on the partner 
anti-quark on the other side of the $C$-periodic boundary. Now we can have a 
single charge in a finite volume. However, $C$-periodicity does break the 
$\Z(3)$ center symmetry \cite{Wie92}, as we see from
\begin{equation}
g(\vec{x},t)^* = g(\vec{x},t + \beta)^* z = g(\vec{x} + L_z \vec{e}_z,
t + \beta) z = g(\vec{x} + L_z \vec{e}_z,t) z^2 = g(\vec{x},t)^* z^2.
\end{equation}
This forces $z^2 = 1$ and as $z \in \Z(3)$, therefore $z = 1$. This means
that we no longer have the freedom to twist the gauge transformations. This
explicit breaking of the $\Z(3)$ center symmetry vanishes in the infinite
volume limit. In a finite volume, $\langle \Phi \rangle$ is always non-zero.
If the quark is confined, then $\langle \Phi \rangle$ vanishes in the infinite
volume limit. Alternatively, if the quark is deconfined, $\langle \Phi \rangle$
remains non-zero, if the infinite volume limit is taken in cubic volumes.

We examine analytically how the Polyakov loop behaves in finite
cylindrical volumes of dimension $L_x \times L_y \times L_z$, where 
$L_z \gg L_x, L_y$. We impose ordinary periodicity in the $x$- and 
$y$-directions and $C$-periodicity in the $z$-direction. First, we consider 
the system at low temperature, where quarks are confined. We  
view the partition function of the system diagramatically as
\begin{equation}
Z = \begin{picture}(90,15)
\put(0,-4){\line(1,0){90}} \put(0,11){\line(1,0){90}}
\put(0,-4){\line(0,1){15}} \put(90,-4){\line(0,1){15}} \put(43,0){$c$}
\end{picture} \ = \exp(- \beta f_c A L_z),
\end{equation}
where $A = L_x L_y$ is the cross-sectional area of the cylinder and $f_c$
is the temperature-dependent bulk free energy density of the confined phase.
The partition function is the sum of the Boltzmann weights over all possible 
configurations of the system. At low temperature, the entire volume is filled 
with bulk confined phase. We obtain $\langle \Phi \rangle$ diagramatically 
from
\begin{equation}
Z \langle \Phi \rangle = \begin{picture}(90,15)
\put(0,-4){\line(1,0){90}} \put(0,11){\line(1,0){90}}
\put(0,-4){\line(0,1){15}} \put(90,-4){\line(0,1){15}} \put(43,0){$c$}
\put(0,3.5){\line(1,0){38}} \put(52,3.5){\line(1,0){38}}
\end{picture} \ = \exp(- \beta f_c A L_z) \Sigma_0 \exp(- \beta \sigma L_z).
\end{equation}
Because the Polyakov loop represents a static test quark and the quark
is in confined bulk phase, there is a QCD string carrying the color flux from 
the quark to its partner anti-quark a distance $L_z$ away on the other side 
of the $C$-periodic boundary. The line in the diagram represents the QCD 
string, with its string tension $\sigma$. Knowing $\langle \Phi \rangle =
\Sigma_0 \exp(-\beta \sigma L_z)$, we find the free energy of the quark is 
$F = -1/\beta \ln \Sigma_0 + \sigma L_z$. This diverges as $L_z \rightarrow 
\infty$, i.e. as the quark and anti-quark are pulled infinitely far apart 
--- the quark is indeed confined.

It has been shown that $SU(3)$ Yang-Mills theory has a first order phase
transition at a temperature $T_c$ \cite{Gav89}. Below $T_c$, external static
quarks are confined. Above $T_c$, quarks are deconfined. 
Deconfinement is signalled by a non-zero value for $\langle \Phi \rangle$  
which breaks the global $\Z(3)$ center 
symmetry. In the deconfined bulk phase, $\langle \Phi \rangle$ can take any 
one of three values, 
\begin{equation}
\Phi^{(1)} = (\Phi_0,0), \ 
\Phi^{(2)} = (- \frac{1}{2} \Phi_0,\frac{\sqrt{3}}{2} \Phi_0), \
\Phi^{(3)} = (- \frac{1}{2} \Phi_0,- \frac{\sqrt{3}}{2} \Phi_0).
\end{equation}
Each value represents a distinct phase that the system can fall into and these
phases can be rotated into one another by an element of $\Z(3)$. Above $T_c$,
the center symmetry is broken, giving us three distinct degenerate deconfined
bulk phases. Because there is more than one phase, different parts of the 
cylinder can be in different deconfined phases. In a cylinder, the phases tend
to be separated by interfaces which are transverse to the long direction. The 
phases on either side of an interface are labelled by their value of $\langle 
\Phi \rangle$, $d_i$ means deconfined phase with the expectation 
value $\langle \Phi \rangle = \Phi^{(i)}$. An interface between two different 
deconfined phases has a temperature-dependent interface tension, which is the
energy cost per unit area $\alpha_{dd} = F/A$. Our diagrammatic expansion of 
the partition function corresponds to summing the Boltzmann weights over all 
configurations with all possible numbers of interfaces, assuming a dilute gas 
of interfaces \cite{Gro93}. The diagrammatic expansion of the partition 
function is
\begin{equation}
Z = \begin{picture}(90,15)
\put(0,-4){\line(1,0){90}} \put(0,11){\line(1,0){90}}
\put(0,-4){\line(0,1){15}} \put(90,-4){\line(0,1){15}} \put(40,0){$d_1$}
\end{picture} \ + \
\begin{picture}(90,15)
\put(0,-4){\line(1,0){90}} \put(0,11){\line(1,0){90}}
\put(0,-4){\line(0,1){15}} \put(90,-4){\line(0,1){15}} 
\put(45,-4){\line(0,1){15}} \put(18,0){$d_2$} \put(63,0){$d_3$} 
\end{picture} \ + \
\begin{picture}(90,15)
\put(0,-4){\line(1,0){90}} \put(0,11){\line(1,0){90}}
\put(0,-4){\line(0,1){15}} \put(90,-4){\line(0,1){15}}
\put(45,-4){\line(0,1){15}} \put(18,0){$d_3$} \put(63,0){$d_2$}
\end{picture} \ + ...
\end{equation}
$C$-periodicity in the $z$-direction means that certain configurations are
not possible, for example, the volume with no interfaces cannot be entirely
filled with either deconfined phase $d_2$ or $d_3$. The terms corresponding 
to the diagrams are
\newpage
\begin{eqnarray}
Z&=&\exp(- \beta f_d A L_z) \nonumber \\
&+& 2 \int_0^{L_z} dz \ \exp(- \beta f_d A z)
\gamma \exp(- \beta \alpha_{dd} A) \exp(- \beta f_d A (L_z - z)) + ... 
\nonumber \\
&=&\exp(- \beta f_d A L_z)
[1 + 2 \gamma \exp(- \beta \alpha_{dd} A) L_z + ...].
\end{eqnarray}
As well as including a Boltzmann weight for the energy cost of the deconfined
phase with free energy density $f_d$, we must also include the additional 
energy cost of an interface. The factor $\gamma$ is included because
the interfaces are not rigid and actually fluctuate. It has been shown that,
in three dimensions, $\gamma$ is to leading order independent of the area
$A$ \cite{Pri83}. To sum over all possible configurations, we integrate
over all possible positions of the interfaces. We sum the interface 
expansion to all orders and find
\begin{equation}
Z = \exp(- \beta f_d A L_z + 2 \gamma \exp(- \beta \alpha_{dd} A) L_z).
\end{equation}
To calculate $\langle \Phi \rangle$, we again sum over all possible
configurations. With the Boltzmann weight of each possible configuration, we
also include the value of the Polyakov loop for that configuration. This
gives the expansion
\begin{eqnarray}
Z \langle \Phi \rangle&=&\exp(- \beta f_d A L_z)
\{\Phi^{(1)} + \int_0^{L_z} dz \ \gamma \exp(- \beta \alpha_{dd} A) \nonumber 
\\
&\times&\frac{1}{L_z}
[\Phi^{(2)} z + \Phi^{(3)}(L_z - z) + \Phi^{(3)} z + \Phi^{(2)}(L_z - z)]
+ ...\} \nonumber \\
&=&\Phi_0 \exp(- \beta f_d A L_z)
\{1 - \gamma \exp(- \beta \alpha_{dd} A) L_z + ...\} \nonumber \\
&=& \Phi_0 \exp(- \beta f_d A L_z - \gamma \exp(- \beta \alpha_{dd} A) L_z),
\end{eqnarray}
where, in the final line, we have summed the expansion to all orders. Dividing
out the partition function, this gives us the result that 
\begin{equation}
\langle \Phi \rangle = \Phi_0 \exp(- 3 \gamma \exp(- \beta \alpha_{dd} A) L_z).
\end{equation}
This is an unusual result --- it means that the free energy of a static quark
is $F = (-1/\beta) \ln \Phi_0 + (3 \gamma/\beta) \exp(- \beta \alpha_{dd} A) 
L_z$. If $A$ is held fixed, this diverges as $L_z \rightarrow \infty$. As the 
quark and anti-quark are
pulled infinitely far apart, the free energy cost becomes infinite. The quark 
is confined, even though there is no QCD string and we only have deconfined 
phases present. The confinement is due to the multiple degenerate 
deconfined phases lining up along the cylinder in definite domains, which
are separated by domain walls. We would not see this effect if we had 
considered a cubic volume, as it would have been entirely filled with
deconfined phase $d_1$. Also, if we had chosen all spatial directions to be
$C$-periodic, the volume would again only contain phase $d_1$. The free energy
$F$ would then be independent of the volume, as we would expect in the 
deconfined phase. This unexpected confinement has an effective ``string 
tension'' $\sigma' = (3 \gamma/\beta) \exp(- \beta \alpha_{dd} A)$ which is 
area and temperature dependent. We will observe this unusual confinement 
numerically and will also use the dependence of $\sigma'$ on the area of the 
cylinder to extract the interface tension $\alpha_{dd}$.

We now derive a new analytical result for the finite volume dependence
of the Polyakov loop near the phase transition. Because the phase transition
is of first order, as we approach the critical temperature $T_c$ from 
above, the three deconfined phases can coexist with the single confined 
phase. We now implement a different choice of spatial boundary conditions. We 
impose $C$-periodicity in all spatial directions. This means that the complex
deconfined phases $d_2$ and $d_3$ do not appear in the cylinder. With these
new spatial boundary conditions, the diagrammatic expansion for the partition 
function near criticality is
\begin{eqnarray}
Z&=&\begin{picture}(90,15)
\put(0,-4){\line(1,0){90}} \put(0,11){\line(1,0){90}}
\put(0,-4){\line(0,1){15}} \put(90,-4){\line(0,1){15}} \put(43,0){$c$}
\end{picture} \ + \
\begin{picture}(90,15)
\put(0,-4){\line(1,0){90}} \put(0,11){\line(1,0){90}}
\put(0,-4){\line(0,1){15}} \put(90,-4){\line(0,1){15}} \put(40,0){$d_1$}
\end{picture} \nonumber \\ \ \nonumber \\
&+&\begin{picture}(90,15)
\put(0,-4){\line(1,0){90}} \put(0,11){\line(1,0){90}}
\put(0,-4){\line(0,1){15}} \put(90,-4){\line(0,1){15}}
\put(30,-4){\line(0,1){15}} \put(60,-4){\line(0,1){15}}
\put(13,0){$c$} \put(40,0){$d_1$} \put(73,0){$c$} 
\end{picture} \ + \
\begin{picture}(90,15)
\put(0,-4){\line(1,0){90}} \put(0,11){\line(1,0){90}}
\put(0,-4){\line(0,1){15}} \put(90,-4){\line(0,1){15}}
\put(30,-4){\line(0,1){15}} \put(60,-4){\line(0,1){15}}
\put(10,0){$d_1$} \put(43,0){$c$} \put(70,0){$d_1$}
\end{picture} + ...
\end{eqnarray}
There are now only interfaces between confined and deconfined phases, which
have an interface tension $\alpha_{cd}$ and a factor $\delta$ which includes 
the fluctuations of these domain walls. We sum the Boltzmann weights for all
possible configurations to obtain
\begin{equation}
Z = 2 \exp(- \beta (f_c + f_d) A L_z/2) \cosh (L_z \sqrt{x^2 + 
\left.\delta'\right.^2}).
\end{equation}
The quantity $x = \beta (f_c - f_d) A/2$ measures how far away we are from
the finite volume critical point (where $f_c = f_d$) and $\delta' = \delta 
\exp(- \beta \alpha_{cd} A)$. To calculate $Z \langle \Phi \rangle$, as
before, we include the value of the Polyakov loop with the Boltzmann weight
for each possible configuration. With $C$-periodicity in all spatial
directions, the static test quark has a neighboring anti-quark in all
directions. If the test quark is placed in a region of confined bulk phase, a
QCD string connects the quark to the nearest anti-quark, which will be in the
$x$- or $y$-direction. Including the additional energy cost of the QCD
string, we calculate
\begin{eqnarray}
\langle \Phi \rangle &=& 
\Big\{ \big[ \frac{\Phi_0 + \Sigma'_0}{2} + 
\frac{x(\Phi_0 - \Sigma'_0)}
{2\sqrt{x^2 + \left.\delta'\right.^2}} \big] 
\exp(L_z \sqrt{x^2 + \left.\delta'\right.^2}) + \\
&& \big[ \frac{\Phi_0 + \Sigma'_0}{2} - 
\frac{x(\Phi_0 - \Sigma'_0)}{2\sqrt{x^2 
+ \left.\delta'\right.^2}} \big] \exp(- L_z \sqrt{x^2 + 
\left.\delta'\right.^2}) \Big\} / 
2\cosh(L_z \sqrt{x^2 + \left.\delta'\right.^2}), \nonumber 
\end{eqnarray}
where $\Sigma'_0 = \Sigma_0 \exp(- \beta \sigma L_{\bot}), 
L_{\bot} = \mbox{min}(L_x, L_y)$. We 
test this new result in two limits. If we are on the deconfined side of the
phase transition, then $f_d < f_c$ and we find that, as the volume grows 
large, $\langle \Phi \rangle = \Phi_0$. This is as we expect --- a domain 
wall costs more and more energy, so the whole volume is filled with deconfined
phase $d_1$ and the quark energy is independent of the volume size. If we are 
on the confined side of the phase transition with $f_d > f_c$, as the volume
becomes large, $\langle \Phi \rangle = \Sigma_0 \exp(- \beta \sigma L_{\bot})$
and $F = (-1/\beta) \ln \Sigma_0 + \sigma L_{\bot}$. The entire volume is
filled with confined phase and the quark is connected to the nearest 
anti-quark by a QCD string. This analytical result shows how the finite volume
behavior of the Polyakov loop changes across the deconfinement phase 
transition with $C$-periodicity in all directions. We also use this 
expression to extract the confined-deconfined interface tension $\alpha_{cd}$.

\section{Numerical investigation of the Potts Model}
From the analytical calculations for $SU(3)$ Yang-Mills theory, we expect to 
see unusual confinement even deep in the deconfined phase. If we can observe 
such behavior numerically, we can extract properties of domain walls, such as 
the deconfined-deconfined interface tension $\alpha_{dd}$. The analytical 
calculation assumes that there are definite deconfined domains which line up 
along the cylinder separated by interfaces. While this is certainly true if 
the volume is arbitrarily large, it is not at all clear if this phenomenon can
be observed in moderately sized volumes. The Polyakov loop expectation value 
should become exponentially small as the cylinder length grows. It will 
certainly be very difficult to detect numerically such a small signal for 
Yang-Mills theory. We want to know if this volume dependence is at all 
observable. We will examine the 3-d 3-state Potts model, which is an effective
model for non-zero temperature $SU(3)$ Yang-Mills theory. We will explain why 
it is an effective model and show how we can observe the unusual confinement 
even in relatively small volumes using highly efficient numerical techniques.

In the 3-d 3-state Potts model, there are spins $\Phi_x$ at positions $x$ which 
can take one of three values $\Phi_x \in \Z(3) = \{ \exp(2 \pi i n/3), n 
= 1,2,3 \}$. The action is
\begin{equation}
S[\Phi] = - \beta \sum_{\langle xy \rangle} \delta_{\Phi_x,\Phi_y},
\end{equation}
where the sum is over nearest neighbors. If we globally rotate all spins by 
$\Phi^{\prime}_x = \Phi_x z, z \in \Z(3)$, the action is unchanged. This 
global symmetry is exactly analogous to the $\Z(3)$ center symmetry of the 
Yang-Mills theory. Numerical work shows that this model has a first 
order phase transition at a non-zero temperature $T_c$ \cite{Jan97}. For 
$T > T_c$, the system is in a disordered phase, where $\langle \Phi 
\rangle$ is zero and the global $\Z(3)$ symmetry is intact. For 
$T < T_c$, there are three ordered phases, where $\langle \Phi \rangle 
\propto \exp(2 \pi i n/3), n = 1,2,3$ and the global $\Z(3)$ symmetry is 
broken. The disordered phase is exactly analogous to the confined phase 
of the Yang-Mills theory, while the three ordered phases correspond to the 
three deconfined phases. Because of this shared global symmetry and phase
structure, the Potts model is an effective theory to describe the
deconfinement phase transition. We look at the behavior of 
Potts spins in cylindrical volumes which are periodic in $x$ and $y$ and
$C$-periodic in $z$. We can again apply $C$-periodicity in the 
$z$-direction because charge conjugation is a symmetry of the Potts 
action. Using the dilute gas of interfaces, we calculate that the average spin 
$\langle \Phi \rangle$ in this volume deep in the ordered (deconfined) 
regime is
\begin{equation}
\langle \Phi \rangle = \Phi_0 \exp(- 3 \gamma \exp(- \beta \alpha_{dd} A) L_z).
\end{equation}
Now we have an interface tension $\alpha_{dd}$ for an interface between two
ordered phases and $\gamma$ accounts for the fluctuation of these
non-rigid domain walls. Previously, the Polyakov loop fell off exponentially
with the cylinder length, signalling confinement. This now becomes the 
exponential fall off of the average spin. Similarly, we calculate the average 
spin near the phase transition applying $C$-periodicity in all spatial 
directions and we see that this depends on $\alpha_{cd}$, the energy cost of an
interface between an ordered and a disordered phase. The analytical results
are identical, translating deconfined and confined bulk phases into 
ordered and disordered bulk phases. 

We measure the average spin $\langle \Phi \rangle$ using a cluster
algorithm \cite{Swe87}. We pick a spin at random from the cylinder as the 
first element of a cluster. Next, we examine all nearest neighbors of this 
spin. If a neighbor has the same spin value, we include it in the cluster 
with probability $P = 1 - \exp(-\beta)$. If a neighboring spin has a 
different value, we do not include it --- all the spins in a cluster have 
the same value. We continue to build a cluster until all neighbors of all 
spins in the cluster have  been examined. In a multi-cluster algorithm, we
keep building clusters until the entire volume is decomposed into many
clusters. Once all the clusters are built, all the spins in 
each cluster are flipped to a new value. However, it is not always possible
to flip a cluster. If a cluster winds around a $C$-periodic direction of
the cylinder an odd number of times, it must contain spins with real value
1. Otherwise, it would not satisfy $C$-periodicity in this direction. The
spins in such a cluster cannot be flipped to a new complex value as this 
would break the $C$-periodicity. Such a cluster is called a wrapping
cluster. Any other type of cluster is called non-wrapping and can always
be flipped. In a multi-cluster algorithm, the average spin is 
\begin{equation}
\langle \Phi \rangle = \frac{1}{N_V} \langle \hspace{0.05in} \sum_x \Phi_x 
\hspace{0.05in} \rangle = \frac{1}{N_V} \langle \hspace{0.05in} \sum_{C} 
\sum_{x \in C} \Phi_x \hspace{0.05in} \rangle = \frac{1}{N_V} \langle 
\hspace{0.05in} \sum_{C} N_C \langle\!\langle \Phi \rangle\!\rangle _C 
\hspace{0.05in} \rangle,
\end{equation}
i.e. the average spin is the average over all spins in the volume 
divided by the total number of spins $N_V$. We can sum all the spins in two 
steps --- first we sum over all the spins contained in a single cluster 
$C$, then we sum over all the clusters. Allowing the spins in each cluster 
to flip to all possible states, we sum the average spin in each cluster 
$\langle\!\langle \Phi \rangle\!\rangle _C$ times the number of spins in each 
cluster $N_C$. What is the average spin in a cluster? In a wrapping cluster, 
the spins must always have real value 1 and so these clusters have 
$\langle\!\langle \Phi \rangle\!\rangle _C = 1$. The spins in a
non-wrapping cluster can take on all possible values and these clusters
have $\langle\!\langle \Phi \rangle\!\rangle _C = 0$. The entire volume is 
decomposed into clusters, the number of spins contained in the wrapping 
clusters is counted and the clusters are flipped to new values. This is 
repeated sufficiently often to numerically estimate the average number of 
spins contained in wrapping clusters. This is an improved estimator because we
only add non-negative numbers in our estimate of $\langle \Phi \rangle$. We 
can also use a single cluster algorithm. Instead of calculating
the average spin by summing the average spin of all clusters, we only
determine the average spin in one single cluster $C$. The probability 
of choosing this cluster from all of them is proportional to its size, i.e. 
$P_{C} = N_{C}/N_V$. The average spin is then
\begin{equation}
\langle \Phi \rangle = \langle \hspace{0.05in} \frac{N_{C}}{N_V} 
\frac{\langle\!\langle \Phi \rangle\!\rangle _C}{P_C} \hspace{0.05in} 
\rangle= \langle \hspace{0.05in} \langle\!\langle \Phi \rangle\!\rangle _C 
\hspace{0.05in} \rangle = P_{\mbox{\small wrap}},
\end{equation}
the sum over all cluster averages is replaced by one cluster average divided
by the probability $P_{C}$ of choosing that cluster. The average
spin in the entire volume equals the average spin in the single cluster
that we build, which equals the probability $P_{\mbox{\small wrap}}$
that the single cluster is a wrapping cluster. To estimate this
probability numerically, we build a single cluster; if it is wrapping, we
count a 1, if it is non-wrapping, we count a 0. We repeat this procedure 
until we can accurately estimate the wrapping probability. Again, this is an 
improved estimator because we only add non-negative numbers in making our 
estimate of $P_{\mbox{\small wrap}}$. For very long and thin cylinders, 
$P_{\mbox{\small wrap}}$ is very small and such a small signal would be 
very difficult to detect if we had positive and negative contributions to our 
estimator with lots of cancellation. The signal is small but the improved 
estimator does the best possible job of numerically estimating the 
probability. A previous study of the Ising model has used an algorithm where 
boundary conditions can be changed depending on the properties of  clusters
\cite{Cas96}.

\begin{figure}
\psfig{figure=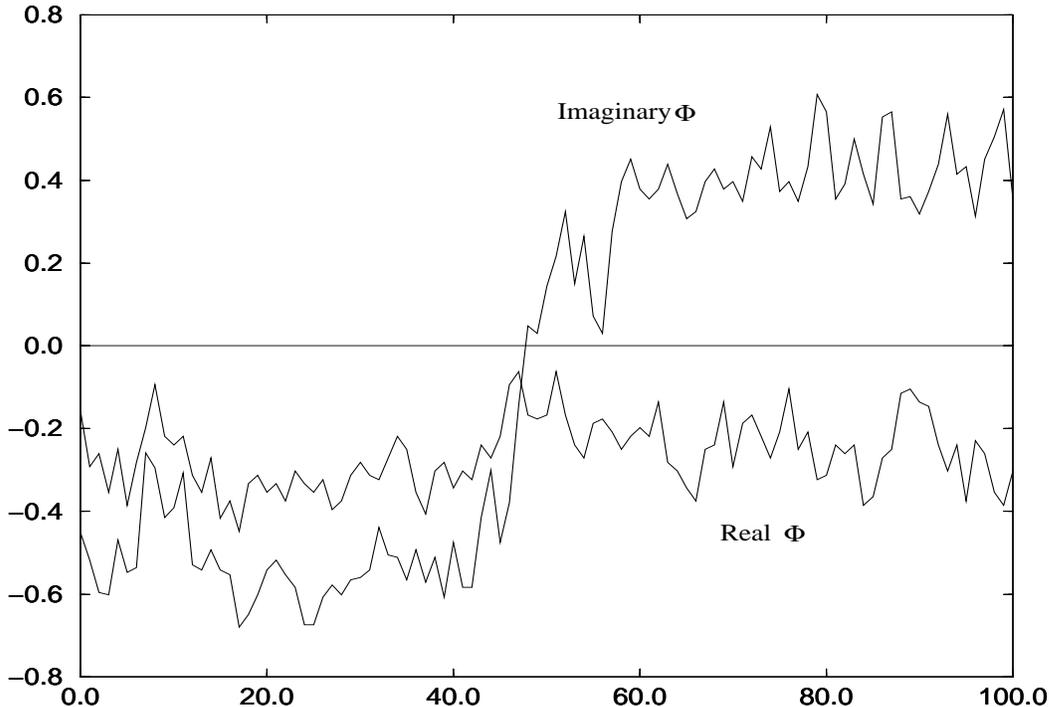,height=5.5in,width=3.7in,angle=270}
\caption{\it The spins $\Phi_x$ are averaged over the transverse plane located
at longitudinal coordinate $z$. The plot is of the real and imaginary parts of 
the planar average of the spin versus $z$. There is one interface, separating 
two of the ordered bulk phases.}
\end{figure}

We first consider the case deep in the ordered phase of the theory, where we
expect $\langle \Phi \rangle$ to decrease exponentially with the cylinder 
length, which is the signal in this effective model for the unusual 
confinement. Here, we use the single cluster algorithm to estimate $\langle 
\Phi \rangle$. Previous work has determined that the phase transition occurs 
at $\beta_c = 0.550565(10)$ \cite{Jan97}. The ordered regime is $\beta > 
\beta_c$. If we go too far away from $\beta_c$, the energy cost of even one 
interface will be very large, an interface will practically never appear in 
the volume and our analytical calculation will not apply. We work in a region
$\beta_c < \beta < 1.005 \beta_c$, where typically there are interfaces 
in the cylinder. The volumes we examine are of transverse dimension 
$L_x = L_y = 12,13,14$ and $L_z$ in the range 80 to 120. In Figure 1, we 
show a typical numerical configuration of a volume of dimension 
$12 \times 12 \times 100$ at a temperature $\beta = 1.003 \beta_c$. This 
particular configuration has one interface with bulk ordered phase $\Phi = 
(-1/2 - i \sqrt{3}/2) \Phi_0$ on the left hand side and ordered phase 
$\Phi = (-1/2 + i \sqrt{3}/2) \Phi_0$ on the right.

\begin{figure}
\psfig{figure=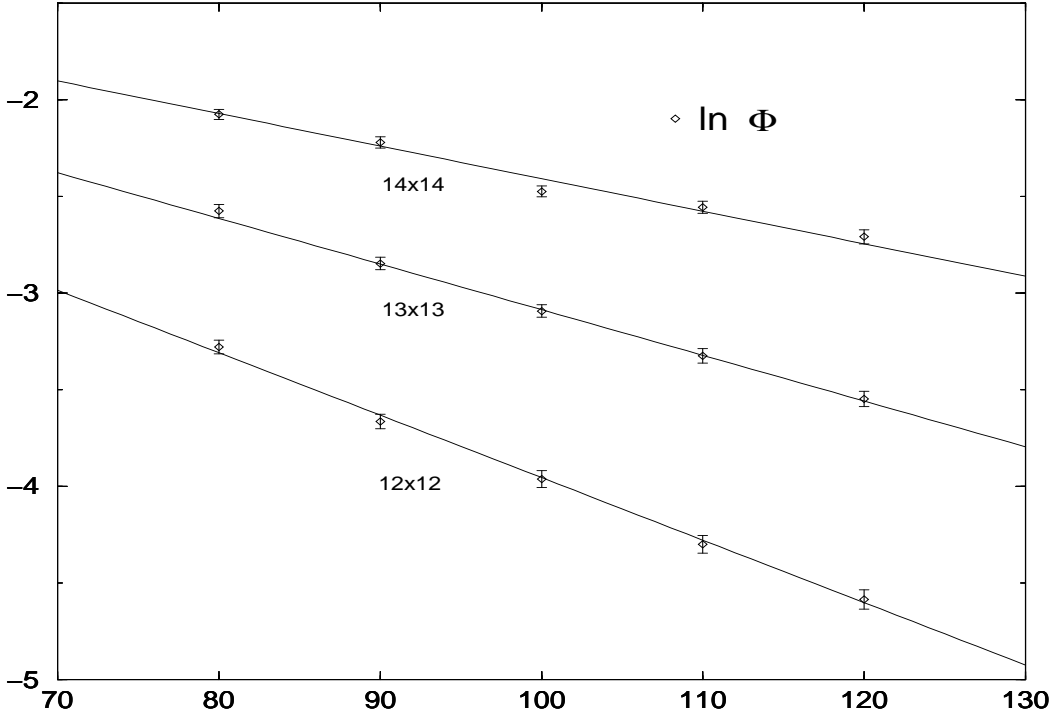,height=5.5in,width=3.7in,angle=270}
\caption{\it A plot of $\ln \langle \Phi \rangle$ versus $L_z$, the 
cylinder length. The lines are the fit to the analytical calculation of 
$\ln \langle \Phi \rangle$ in the ordered phase regime.}
\end{figure}
 

\begin{figure}
\psfig{figure=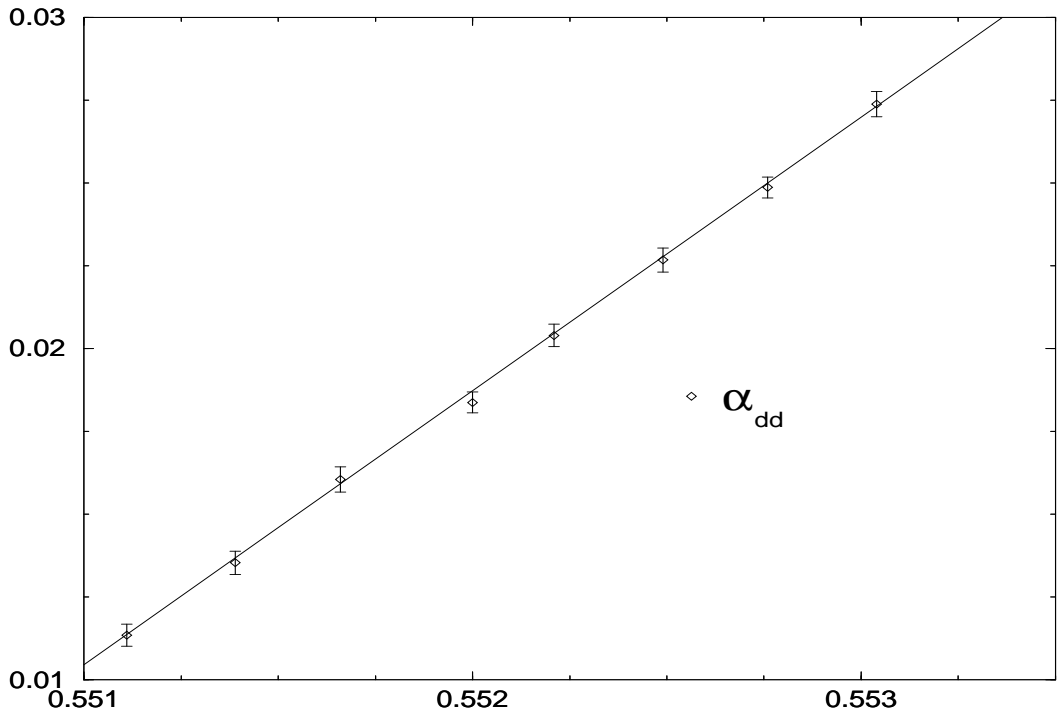,height=5.5in,width=3.7in,angle=270}
\caption{\it The ordered-ordered interface tension $\alpha_{dd}$ as a
function of $\beta$. The fit is a straight line.}
\end{figure}

In Figure 2, we have plotted the numerical results for $\ln \langle \Phi 
\rangle$ in cylinders of various lengths and areas obtained at a 
particular temperature, here $\beta = 1.0035 \beta_c$. We see clearly that
the average spin falls off exponentially with the length of the cylinder. The 
single cluster algorithm and associated improved estimator are able to 
detect this extremely small signal. We see that our analytical calculation, 
which is certainly true for arbitrarily large volumes, also holds for volumes 
of moderate size. We fit the numerical data to the analytical formula and 
each line in Figure 2 corresponds to cylinders with a fixed area, i.e. 
$12 \times 12, 13 \times 13$ and $14 \times 14$. All three lines are obtained 
from one fit, which gives the numerical value of $\alpha_{dd}(\beta)$ with 
$\chi^2$ per degree of freedom $\sim 0.7$. We take this to be an indication of
a good fit. We have performed simulations at many values of $\beta$, including
one previously examined in the literature, allowing us to make a direct 
comparison of the interface tension. At $\beta = 0.552$, a value of 
$\alpha_{dd} = 0.01796(14)$ has previously been determined \cite{Pro94}. Using
our technique, we find $\alpha_{dd} = 0.01838(42)$, which is in agreement 
within error bars. In Figure 3, we have plotted the values we have determined 
for $\alpha_{dd}$ as a function of $\beta$. The data fits very well to a 
straight line, with $\chi^2/$d.o.f $\sim 0.2$. The fit is
\begin{equation}
\alpha_{dd}(\beta) = -4.54(4) + 8.26(8) \beta,
\end{equation}
where the uncertainties in the final digits are paranthesized. The 
uncertainties appear rather large, but are actually highly correlated. At the 
critical point, it is expected that $\alpha_{dd}(\beta_c) = 2 \alpha_{cd}$ due
to a phenomenon called complete wetting. It is tempting to extrapolate our 
linear fit to the critical point to estimate $\alpha_{cd}$. However, our 
configurations do not contain disordered phase and our numerical data cannot 
know anything about the energy cost of an ordered-disordered domain wall. We 
cannot extrapolate the fit deep in the ordered phase to the critical point.

\begin{figure}
\psfig{figure=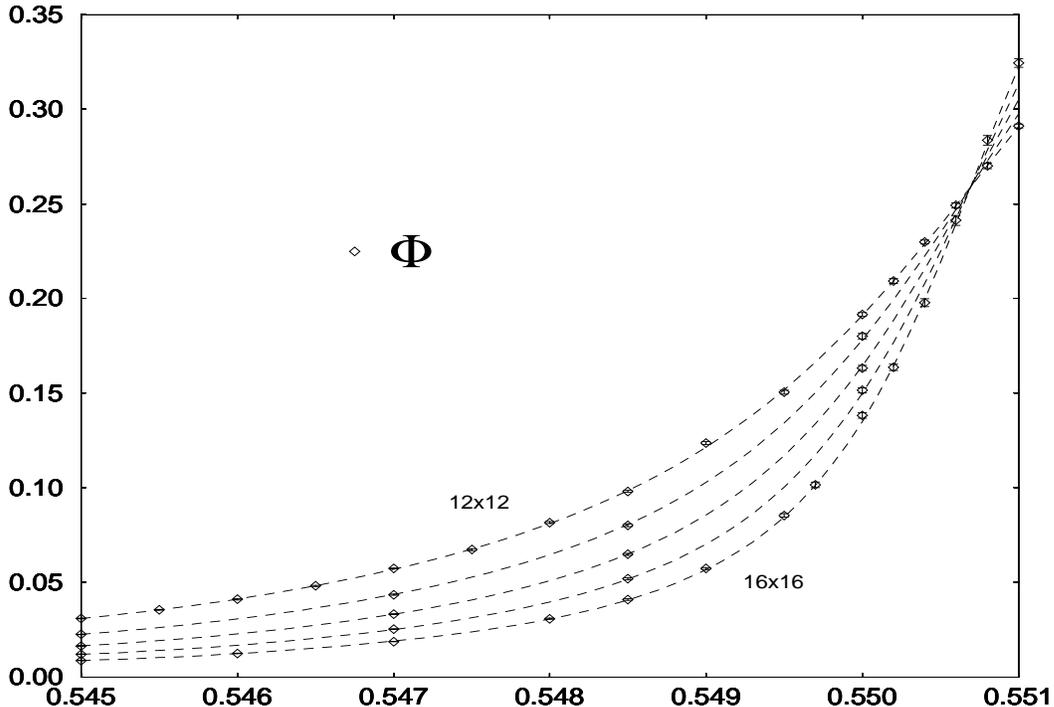,height=5.5in,width=3.7in,angle=270}
\caption{\it A plot of $\langle \Phi \rangle$ versus $\beta$ in cylinders
of length $L_z = 80$ and areas from $12 \times 12$ to $16 \times 16$. The
curves are the fit to the analytical behavior of $\langle \Phi \rangle$
across the phase transition.} 
\end{figure}

We cannot use the single cluster algorithm to verify our new analytical 
calculation near the phase transition, where we use $C$-periodicity in all 
directions. Using the single cluster algorithm, $\langle \Phi \rangle$ is 
equal to $P_{\mbox{\small wrap}}$. With $C$-periodicity in all 
directions, the probability of building a wrapping cluster is much greater and
so the numerical signal is much larger. However, a wrapping cluster cannot be 
flipped to a new state and the configuration is unchanged after this 
iteration. The more often we build a wrapping cluster, the longer it takes for
configurations to become independent of one another. Thus, a larger signal 
also means larger autocorrelations. In this case, the single cluster algorithm
gives much larger integrated autocorrelation times than with
$C$-periodicity in the long direction only. This makes the algorithm 
inefficient and we cannot detect finite volume dependence in $\langle \Phi 
\rangle$. In this instance, the multicluster algorithm is far superior. By 
decomposing the entire volume into clusters, there will always be many 
non-wrapping clusters which can be flipped and configurations very quickly 
become decorrelated. Using the multicluster algorithm, we examine cylinders 
with dimensions $L_x = L_y = 12,13,14,15,16$ and $ L_z= 60,70,80,90$ and
at temperatures in the range $\beta \in [0.545,0.551]$. We fit all of the data
to the predicted analytical behavior of $\langle \Phi \rangle$. In Figure 4, 
we plot some of the numerical data and the fit of $\langle \Phi \rangle$ as a 
function of $\beta$. From the fit of all the data, we find the ordered-disordered 
interface tension is $\alpha_{cd} = 0.0015(2)$, with $\chi^2$/d.o.f $\sim 1.1$. 
Translated into our notation, a value of $\alpha_{cd} = 0.00148(2)$ has previously
been found \cite{Jan97}. From the fit, we also obtain $\beta_c = 0.55070(17)$, in 
comparison to the previously quoted value $\beta_c = 0.550565(10)$ \cite{Jan97}. 
These values are consistent and we feel that the relatively large percentage 
errors in our estimates of the interface tension and the critical temperature may 
be because we work in smaller volumes. There are also other techniques to extract 
the interface tension near criticality \cite{Sch94}. 

\section{Conclusions}
Analytically, we predicted that a very unusual type of confinement occurs even
in the deconfined phase of $SU(3)$ Yang-Mills theory. The free energy of a 
single static quark diverges, even though it is contained in deconfined bulk
phase. The signal for this unusual confinement is that the expectation value
of the Polyakov loop becomes exponentially small. We have successfully 
observed this phenomenon numerically in an effective theory. It was not a 
priori clear that the exponentially small signal could be measured or that 
this confinement would be observable in moderately sized volumes. The single 
cluster algorithm and associated improved estimator can detect this very small
signal in reasonably sized cylinders and accurately determine the energy cost 
of domain walls between distinct bulk phases. Our new analytical result 
predicts the finite volume behavior of the Polyakov loop across the 
deconfinement phase transition and we observe this behavior numerically, 
allowing us to estimate the energy cost of domain walls at criticality. 
Interestingly, this is an example where a multicluster algorithm is far 
superior to a single cluster algorithm.

\section{Acknowledgements}
I would like to acknowledge the essential contribution of U.-J. Wiese
to this project.

\end{document}